\documentstyle[12pt,epsf]{article}

\let\d=\delta
\let\g=\gamma

\let\r=\rho
\let\s=\sigma

\newcommand{\beq}{\begin{equation}}
\newcommand{\eeq}{\end{equation}}
\newcommand{\bea}{\begin{eqnarray}}
\newcommand{\eea}{\end{eqnarray}}

\newcommand{\Dsl}[1]{\slash\hskip -0.20 cm #1}

\newcommand{\nbox}{{\,\lower0.9pt\vbox{\hrule \hbox{\vrule height 0.2 cm \hskip
0.2 cm \vrule height 0.2 cm}\hrule}\,}}

\DeclareFixedFont{\xiiss}{OT1}{cmss}{m}{n}{12}
\DeclareFixedFont{\ixss}{OT1}{cmss}{m}{n}{9}
\DeclareFixedFont{\cmrnine}{OT1}{cmr}{m}{n}{9}

\newcommand{\CC}{\hbox{\xiiss C\kern-.4emI}}
\newcommand{\RR}{\hbox{\xiiss R\kern-.45emI}}
\newcommand{\ZZ}{\hbox{\xiiss Z\kern-.4emZ}}
\newcommand{\CCs}{\hbox{\ixss C\kern-.4emI}}
\newcommand{\ZZs}{\hbox{\ixss Z\kern-.4emZ}}
\newcommand{\pa}{\partial}

\newcommand{\pasl}{\pa\kern-.55em /}
\def\href#1#2{#2}

\begin{document}
\begin{titlepage}
\title{ 
        \begin{flushright}
        \begin{small}
        RU-NHETC-2000-10\\
        hep-th/0003227\\
        \end{small}
        \end{flushright}
        \vspace{1.cm}
  Noncommutative  Gauge  Theory,  Divergences and Closed Strings
 }
\author{
Arvind Rajaraman\thanks{e-mail: \tt arvindra@physics.rutgers.edu}
\ and
Moshe Rozali\thanks{e-mail: \tt rozali@physics.rutgers.edu}\\
\\
        \small\it Department of Physics and Astronomy\\
        \small\it Rutgers University\\
        \small\it Piscataway, NJ 08855
}

\maketitle

\begin{abstract}
 We study the renormalization of non-commutative gauge theories with matter. 
 As in the scalar field theory cases, there are logarithmic infrared
divergences resulting from integrating out high momentum modes.
 In order to reproduce the correct infrared behaviour, the Wilsonian effective action has to include certain '`closed string`` modes with prescribed couplings.

In the  case of quiver gauge theories, realized in string theory on orbifolds, we identify the required modes with a set of twisted sector fields. These closed string modes have exactly the prescribed  couplings to  correct the Wilsonian effective action. This provides a concrete origin for  the 
 appearance of closed string modes in noncommutative field theories.

\end{abstract}

\end{titlepage}

\section{Introduction}

  Noncommutative field theories have been the focus of much interest recently. Following their appearance in Matrix theory \cite{cds}, and in string theory
\cite{sw}, there has been renewed interest in their perturbative study. 
 In \cite{msv} scalar field theories have been studied perturbatively. Surprisingly, some remnants of stringy behaviour are visible even in the
perturbation theory. Subsequent discussions of the scalar field theories   include classical solutions \cite{solitons}, and finite temperature effects \cite{temprature}.

In \cite{msv} the Wilsonian effective action of non-commutative scalar theories
 was discussed. The non-commutativity parameter acts as an effective
 ultra-violet cut-off, suppressing all ultraviolet divergences in non-planar
 diagrams. Instead, one encounters curious infrared divergences in the 1PI
 effective action. These divergences come from high momentum integration. 
Therefore they are incorrectly cut-off in the Wilsonian approach.

In order to repair the Wilsonian approach, a general  procedure was 
suggetsted in \cite{msv}.   The Wilsonian effective action includes extra light
(non-propagating) modes, which have the required couplings to correct the infrared behaviour of the Wilsonian effective action. Those modes were interpreted as closed string modes.

 We are interested here in examining the issues raised in \cite{msv} in a context which is more closely related to string theory. To this effect we study the perturation theory 
of supersymmetric non-abelian gauge theory with matter\footnote{Noncommuatative gauge theories 
have  been studied perturbatively in \cite{gauge,susskind}.}.
 One can then discuss noncommutative theories that have both  the above mentioned infrared divergences (unlike the $\mathcal{N}$=4 theory), and also a string theory realization (unlike the scalar field theories).

 We find that similar infrared effects arise also in those theories. In particular the procedure of adding closed string modes to the effective action \cite{msv} works in the present context as well. Studying the stringy realization of the gauge theories reveals the origin of the extra modes. They are indeed closed string modes of the underlying string theory.

The paper is organized as follows:

 In the next section we introduce the classical action for the noncommutative 
theories we study.  These  theories include an arbitrary product gauge group, with unitary factors, coupled to matter in the fundamental and in the adjoint representations. A somewhat surprising result is the existence of gauge invariant
local operators if one includes matter in the fundamental representation. This is unlike the case of the pure gauge theory.

We then turn to studying the renormalization properties of noncommutative theories. After reviewing the results in \cite{msv} about the renormalization of scalar theories, we calculate similar results in the gauge theory case. We concentrate on the  $\beta$ functions of each of the gauge factors, and on the associated IR divergences.

As was the case in \cite{msv}, we find that the UV does not decouple from the IR physics. When forming a Wilsonian effective action, there is a need to add
some ''closed string'' modes to account for infrared divergences. We use the procedure outlined in \cite{msv} in the present context to write explicitly the required modes and their couplings.

 In the rest of the paper we study a string realization of such theories. A general  class of 
$\mathcal{N}$=1 supersymmetric theories, the so called quiver theories, can be realized as the worldvolume 
theories of branes transverse to an orbifold singularity. We review the construction of the quiver gauge theories and specify their matter content. 
The limit considered by Seiberg and Witten \cite{sw} should then yield a noncommutative version of the quiver gauge theories.

In the last section we discuss the stringy realization of the closed string modes required to fix the Wilsonian effective action. We conclude that the 
twist fields are indeed of the right form to be these closed string modes. Their inclusion in the effective action summarizes the effect of the high momentum 
gauge theory modes that have been integrated out. They have the correct couplings by virtue of a relation between the $\beta$ function coefficients and
twist field tadpoles, studied in \cite{lr}. We identify all the required modes 
in the large $N_c$ limit of the gauge theories, and point out a universal 
discrepency to do with the overall $U(1)$ factor. 

 We conclude by discussing open questions regarding the (absence of)
massive closed string contributions to the infrared divergences, and  quadratic divergences in the orbifold realization of non-supersymmetric quiver gauge theories.

The relation between the perturbative calculation of \cite{msv} and string theory has been also discussed in \cite{string}.

\section{Noncommutative Gauge Theories with~Matter}

 In a ordinary non-abelian gauge theory matter fields transform by a matrix representation of the gauge group. For a non-commutative gauge theory there can be two types of representations: left modules and right modules. This simply 
asserts that the  gauge group acts on the field from the left or from the right.  Gauge invariance restricts  possible couplings of such matter fields as described below. 

 Suppose $A_\mu$ is a non-commutative gauge field, transforming in the adjoint representation of $G= U(N)$. The gauge transformation of $A_\mu$ is:

\beq
\delta A_\mu = - \partial_\mu \epsilon + i\epsilon * A_\mu
-i A_\mu * \epsilon 
\eeq

 Where we suppress the fundamental $U(N)$ indices $i,j= 1,...,N$ in the gauge field $A_\mu$ and in the gauge parameter $\epsilon$.  With respect to the global part of the gauge transformation, the gauge field transforms as a bi-module: $G$ acts simoultaneously
from the left and from the right. The field strength that transforms covariantly is defined as:

\beq
F_{\mu\nu} = \partial_\mu A_\nu - \partial_\nu A_\mu  +i A_\mu * A_\nu
-i A_\nu  * A_\mu
\eeq

Then one can write the  standard action for the gauge fields:

\beq
I= \frac{1}{4g^2}\int d^4x \, Tr \left[ F_{\mu\nu} * F^{\mu\nu} \right]
\eeq

Raising and lowering of spacetime indices is done with flat space  metric.
 The trace is in the fundamental representation of $U(N)$.
 The kinetic action for several $U(N)$ gauge factors is simply the sum of this action for each of the gauge factors.

We are now ready to discuss matter couplings (matter couplings are discussed in \cite{classical}).
 The gauge transformations for the fundamental  left or right modules are:
\bea
&\delta \Phi_L = i \epsilon \Phi_L \nonumber\\
&\delta \Phi_R = -i \Phi_R \epsilon
\eea

One can define  covariant derivatives which transform similarly, as follows:
\bea
&D_\mu \Phi_L = \partial_\mu \Phi_L  +i A_\mu \Phi_L \nonumber\\
&D_\mu \Phi_R = \partial_\mu \Phi_R  -i  \Phi_R A_\mu 
\eea

In the commutative limit, the left and right modules go over to fields
in the fundamental and anti-fundamental respectively. It is  natural
to define Hermitian conjugation which pairs up left and right modules. In the commutative limit this notion of Hermitian conjugation goes over to the usual one.

A  gauge invariant action can be written for a  field and its Hermitian conjugate, which transform in  the fundamental
left and right modules, respectively.  The gauge invariant kinetic term is:

\beq
 I = \int d^4 x \, Tr \left[ D_\mu \bar{\Phi}_R D^\mu \Phi_L\right]
\eeq

We note that the kinetic term is gauge invariant before integration. The Lagrangian density provides therefore a gauge invariant local operator. 
One can easily consruct other such operators.
 Similarly one can use the fundamental representation to construct Wilson lines, which are gauge invariant for any particular path chosen. This is in contrast to the pure gauge theory case, where no such objects exist.

In the following we are interested in quiver gauge theories. These are product
gauge theories with gauge group  factors $U(N)$. The  matter fields  transform in  bi-fundamental
representation.  In the non-commutative case this means that one factor of the gauge group acts from the left, and another from the right. We denote such a field schematically by $\Phi_{LR}$, and its hermitian conjugate by $\bar{\Phi}_{RL}$. The covariant derivatives of the fields  $\Phi_{LR},\bar{\Phi}_{RL}$  are:

\bea
& D_\mu \Phi = \partial\mu \Phi + i A_\mu^{(1)} \Phi -i \Phi A_\mu^{(2)} \nonumber \\
& D_\mu \bar{\Phi} =\partial\mu \bar{\Phi} + i A_\mu^{(2)} \bar{\Phi} -i \bar
{\Phi} A_\mu^{(1)}
\eea

 A gauge invariant kinetic term is then:

\beq
 I = \int d^4 x \, Tr \left[ D_\mu \bar{\Phi} D^\mu \Phi\right]
\eeq

Note that the Lagrangian density now is gauge invariant with respect to one of the gauge factors. With respct to the other gauge factor acting on $\Phi_{LR}$,  it is gauge invariant only up to total derivative.  Therefore, in a quiver gauge theories it is still  difficult to construct simple gauge invariant local operators.

In addition we note that there is no longer a decoupled $U(1)$ in this case, as
is the case of the single $U(N)$ gauge theory. 

 The quiver gauge theories appear naturally in string theory, as reviewed below.
 In the next section we study The Wilsonian effective action of these gauge theories.  We find appearance of closed string modes similar to \cite{msv}.

\section{IR Divergences in Scalar Field Theory}
We first review IR divergences in noncommutative
$\phi^4$ theory. The action is
\beq
S= \int d^4x \left[ \frac{1}{2} (\partial_\mu \phi)^2 + \frac{1}{2} m^2 \phi^2
+ \frac{1}{4!} g^2 \phi*\phi*\phi*\phi \right]
\eeq

\begin{figure}[!ht]
\centerline{\epsfysize=3cm \epsfbox{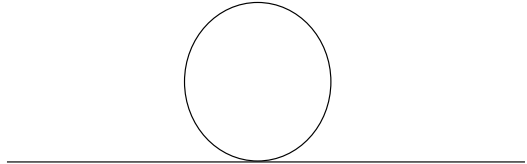}}
\bigskip
\caption{Feynman diagram for the scalar theory}
\end{figure}

This 2 point function is calculated using the diagram above, which yields:
\beq
\Gamma^{(2)} = \frac{g^2}{3(2\pi)^4} \int \frac{d^4k}{k^2 +m^2}cos({k.p\over 2})
\eeq

Rewrite the integrals in terms of
Schwinger parameters using
\beq
\frac{1}{k^2+m^2} = \int d\alpha e^{-\alpha(k^2 + m^2) }
\eeq

The integrals are regulated by multiplying the integrands by
$e^{-\frac{1}{\alpha \Lambda^2}}$. Then
\bea
\Gamma^{(2)}=\Gamma^{(2)} _{planar}+\Gamma^{(2)} _{non-planar}
\\
\Gamma^{(2)} _{planar} = \frac{g^2}{3 (2\pi)^4} \int d\alpha \int {d^4k}
e^{- \alpha(k^2 +m^2)- \frac{1}{\alpha \Lambda^2}}
\\
\Gamma^{(2)} _{non-planar} = \frac{g^2}{6 (2\pi)^4} \int d\alpha \int {d^4k}
e^{- \alpha(k^2 +m^2)- \frac{1}{\alpha \Lambda^2 +ikp}}
\eea

These can evaluated to give 
\bea
\Gamma^{(2)} _{planar} =\frac{g^2}{3 (2\pi)^4}\left( \Lambda^2
-m^2ln\left({\Lambda^2\over m^2}\right)+ ...\right)
\nonumber
\\
\Gamma^{(2)} _{non-planar} = \frac{g^2}{6 (2\pi)^4}
\left( \Lambda_{eff}^2
-m^2ln\left({\Lambda_{eff}^2\over m^2}\right)+ ...\right)
\eea
where 
\bea
\Lambda_{eff}^2= {1\over {1\over \Lambda^2}+\tilde{p}^2}
\nonumber
\\
\tilde{p}_j=p^i(\Theta)_{ij}
\eea

The 1PI effective action is then 
\bea
S= \int d^4p  \frac{1}{2}(p^2+m^2) +  
{g^2\over 96\pi^2( {1\over \Lambda^2}+\tilde{p}^2)} 
-{g^2\over 96\pi^2}ln\left( {1\over M^2( {1\over \Lambda^2}+\tilde{p}^2 )}\right)
\eea

In \cite{msv}, the authors showed that the first new term in the above 1PI action could be obtained from a Wilsonian 
action with an extra $\chi$ field coupled linearly to $\phi$.

\bea
S_{eff}= \int d^4x  \frac{1}{2} (\partial_\mu \phi)^2 + \frac{1}{2} m^2 \phi^2
+ \frac{1}{4!} g^2 \phi*\phi*\phi*\phi
\nonumber
\\
+  \int d^4x  \frac{1}{2}(\partial \chi)o(\partial \chi)+\frac{1}{2}\Lambda^2(\partial o\partial \chi)^2+{i\over \sqrt{96\pi^2}}g\chi\phi
\eea

Integrating out $\chi$ correctly reproduces the first correction to
the 1PI action.

Similarly the logarithmic term can be obtained by
adding a second field $\chi_2$ with a coupling
$\int d^4xg\chi_2\phi$ and a logarithmic propagator
\beq
\langle \chi_2(p) \chi_2(p)\rangle= 
-2 ln\left( {{1\over \Lambda^2}+\tilde{p}^2\over \tilde{p}^2} \right)
\eeq 

The inclusion of fields with logarithmic propagators seems arbitrary,
but \cite{msv} showed that there was a natural interpretation of
these fields as coming from closed string fields living in
2 extra dimensions. The 3+1 dimensional theory where the
$\phi$- quanta live is taken to be a 3-brane living
in 5+1 dimensions. The $\chi_2$ fields live in all 5+1
dimensions, but couple to the $\phi$ fields at the brane location. The $\chi_1$ fields live on the brane only.

The $\chi$ fields have the closed string metric, which is
$g^{\mu\nu}=-{1\over \alpha^{'2}}(\Theta^2)^{\mu\nu}$ in the brane directions, and
$\delta^{\mu\nu}$ in the transverse directions. Furthermore, there
is a 
cutoff ${1\over \alpha'\Lambda}$ on the transverse momenta of the $\chi$ fields.

Then the effective 4-dimensional propagator of the $\chi$ fields is
\bea
\langle \chi_2(p) \chi_2(p)\rangle=\int^{1\over \alpha'\Lambda} {d^2q\over (2\pi)^2} {1\over {\tilde{p}^2\over \alpha^{'2}} +q^2}
\\
={1\over 4\pi} ln\left({{1\over \Lambda^2}+\tilde{p}^2\over \tilde{p}^2}\right)
\eea
as required.

\section{Gauge Theories}

We  start with the case of $\mathcal{N}$=1 $U(N)$ noncommutative gauge theory.

The $U(N)$ gauge field can be written as:
\bea
A_\mu=A^A_\mu T^A= {1\over \sqrt{N}}A_\mu^0 {\bf 1}+ A_\mu^a t^a
\eea
where $t^a$ are $SU(N)$ matrices. 

The  standard action for the gauge fields is:
\beq
I= \frac{1}{4g^2}\int d^4x \, Tr \left[ F_{\mu\nu} * F^{\mu\nu} \right]
\eeq
with
\beq
F_{\mu\nu} = \partial_\mu A_\nu - \partial_\nu A_\mu  +i A_\mu * A_\nu
-i A_\nu * A_\mu
\eeq

In momentum space one can write:
\bea
F_{\mu\nu}=p_\mu A_\nu^AT_A-p_\nu A_\mu^AT_A
+ig( e^{i\tilde{p}^{(1)}p^{(2)}}A_\mu^A(p^{(1)})T_AA_\nu^B(p^{(2)})T_B\nonumber
\\
-e^{i\tilde{p}^{(2)}p^{(1)}}A_\nu^B(p^{(2)})T_BA_\mu^A(p^{(1)})T_A )
\nonumber
\\
=p_\mu A_\nu^AT_A-p_\nu A_\mu^AT_A
+igA_\mu^A(p^{(1)})A_\nu^B(p^{(2)})( cos( \tilde{p}^{(1)}p^{(2)} )[T_A,T_B] \nonumber
\\
+isin(\tilde{p}^{(1)}p^{(2)} )\{ T_A,T_B \} )
\eea

The interaction terms are then
\bea
A_\mu^A(p^{(1)})A_\nu^B(p^{(2)})A^{\nu C}(p^{(3)})
(p^{(1)}_\mu cos( \tilde{p}^{(2)}p^{(3)})tr(T_A [T_B,T_C] +
\nonumber
\\
ip^{(1)}_\mu sin(\tilde{p}^{(1)}p^{(2)} ) T_A\{ T_B,T_C \})
\nonumber
\eea
and 
\bea
A_\mu^A(p^{(1)})A_\nu^B(p^{(2)})A^{\mu C}(p^{(3)})A^{\nu D}(p^{(4)})
\nonumber
\\
( cos( \tilde{p}^{(1)}p^{(2)} )[T_A,T_B] +isin(\tilde{p}^{(1)
}p^{(2)} )\{ T_A,T_B \})
\nonumber\\
( cos( \tilde{p}^{(3)}p^{(4)} )[T_C,T_D] +isin(\tilde{p}^{(3)
}p^{(4)} )\{ T_C,T_D \})
\eea

We wish to compute the 1PI two point function $\langle F F\rangle$ which is obtained
from the diagrams in Fig. 2.

\begin{figure}
\vskip 5cm
$$
\begin{array}{ccc}
\epsfxsize=5cm
\epsfysize=5cm
\epsfbox[200 290 400 440]{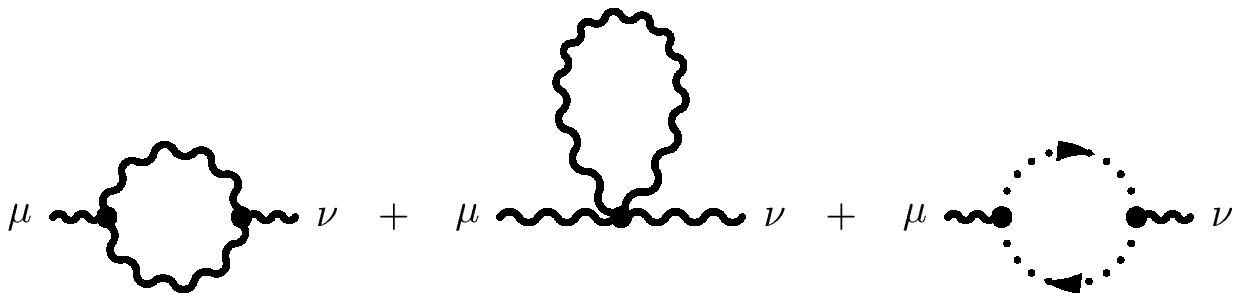} & \epsfxsize=5cm
\epsfbox[200 115 400 265]{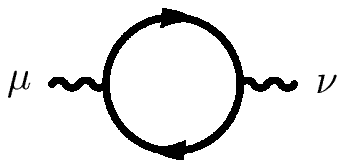}
\end{array}
$$
\vskip -11cm
\caption{Feynman graphs }
\end{figure}


Note that every interaction involving an anticommutator is
down by a factor $\sqrt{N}$ due to the normalization of $A^0$.
We  start by calculating the terms which are leading
order in $N$. To this order, the vertices are identical to
the commutative $SU(N)$ gauge theory with the replacement
\bea
f^{abc}\rightarrow f^{abc}cos( \tilde{p}^{(1)}p^{(2)} )
\eea

The diagrams are each of the form
\bea
\int {d^4p\over (2\pi)^4}{i\over p^2}{i\over (p+q)^2}g^2 C_2(G)\delta^{ab}
cos^2(\tilde{p}q)N^{\mu\nu}
\eea
where
\bea
N^{\mu\nu}_{(1)}={1\over 2}(g^{\mu\rho}(q-p)^\s + g^{\r\s}(2p+q)^\mu+
g^{\s\mu}(-p-2q)^\r)~~~~~~~~~~~\nonumber
\\
(\d^\nu_\r(p-q)_\s+
g_{\r\s}(-2p-q)^\nu+\d^\nu_\s(p+2q)_\r)
\nonumber
\\
N^{\mu\nu}_{(2)}=3(p+q)^2g^{\mu\nu}~~~~~~~~~~~~~~~~~~~~~~~~~~~~~~~~~~~~~~~~~~~~~~~~
\nonumber
\\
N^{\mu\nu}_{(3)}=(p+q)^\mu p^\nu~~~~~~~~~~~~~~~~~~~~~~~~~~~~~~~~~~~~~~~~~~~~~~~~~~~~
\nonumber
\\
N^{\mu\nu}_{(4)}=-tr[ \g^\mu(i\Dsl[k])\g^\nu(i\Dsl[k+q])]~~~~~~~~~~~~~~~~~~~~~~~~~~~~~~
\eea
We can combine the denominators by
\bea
{1\over p^2(p+q)^2}=\int_0^1dx {1\over ((1-x)p^2+x(p+q)^2)^2}
= \int_0^1dx {1\over (P^2-M^2)^2}
\eea
where $P=p+xq$ and $M^2=-x(1-x)q^2$.
We then write the $N^{\mu\nu}_{(i)}$ in terms of $P,q$. We can
drop terms linear in $P$ by symmetry. 

The diagrams are then of the form
\bea
\int_0^1dx \int {d^4P\over (2\pi)^4}{1\over (P^2-M^2)^2}g^2 C_2(G)\delta^{ab}cos^2(\tilde{P}q)\bar{N}^{\mu\nu}
\eea
with
\bea
\bar{N}^{\mu\nu}_{(1)}={1\over 2}(-2g^{\mu\nu}P^2-10P^\mu P^\nu-
g^{\mu\nu}q^2( (2-x)^2+(1+x)^2) 
\nonumber
\\
+q^\mu q^\nu( -2(1-2x)^2+2(1+x)(2-x))
\nonumber
\\
\bar{N}^{\mu\nu}_{(2)}=3g^{\mu\nu}( P^2+(1-x)q^2 )~~~~~~~~~~~~~~~~~~~~~~~~~~~~~
\nonumber
\\
\bar{N}^{\mu\nu}_{(3)}=P^\mu P^\nu-q^\mu q^\nu x(1-x)~~~~~~~~~~~~~~~~~~~~~~~~~~
\nonumber
\\
\bar{N}^{\mu\nu}_{(4)}=4P^\mu P^\nu-2g^{\mu\nu}P^2+2g^{\mu\nu}q^2x(1-x)-2q^\mu q^\nu x(1-x)
\eea

The terms linear in $P^2$ in the above expressions give
quadratic divergences which are cancelled in the usual commutative
case, but give unpleasant IR divergences in the noncommutative
case \cite{susskind}. However, in the supersymmetric case, these divergences cancel.

The terms quadratic in $q$ are then summable to give the final answer
\bea
\int_0^1dx \int {d^4P\over (2\pi)^4}{1\over (P^2-M^2)^2}g^2 C_2(G)\delta^{ab}cos^2(\tilde{P}q)(q^\mu q^\nu-g^{\mu\nu}q^2)
\eea
which is manifestly gauge invariant as in the commutative case.

The noncommutative case therefore differs from the commutative answer only
through the replacement
\bea
\int {d^4P\over (2\pi)^4}{1\over (P^2-M^2)^2}\rightarrow \int {d^4P\over (2\pi)^4}{1\over (P^2-M^2)^2}cos^2(\tilde{P}q)
\eea
and this is always the case if the quadratic divergences cancel in the integrand. As
shown in \cite{susskind}, this is always the case in supersymmetric theories.
The LHS of the equation above produces logarithmic UV divergences
(in the commutative theory). The RHS (in the noncommutative theory)
produces logarithmic IR divergences. The above analysis
says that we can obtain the logarithmic IR divergences
of the noncommutative theory by replacing
$ln \Lambda$ of the commutative theory by
$ln \Lambda_{eff}$. This is identical to the results in \cite{msv}
for the scalar field theory case.

Now the effective action of the usual Yang-Mills theory is of the
form
\bea
S=\int Tr F^2 + \beta(g) ln({p^2\over \Lambda^2}) Tr F^2+\cdots
\eea
where the second term is from the running coupling.

The noncommutative effective action is then of the form
\bea
S=\int Tr F^2 + \beta(g) ln( \tilde{p}^2) Tr F^2+\cdots
\eea
Thus, the coefficient of the IR divergence is proportional to the beta function
coefficient, to the leading order in $N$. This is also the case for the matter diagrams.
Hence, the beta functions for the quiver gauge theories are accompanied
by IR divergences with the same coefficient.
This will be important in the relation to closed strings.

For completeness, we calculate the subleading ${1\over N}$ 
corrections to the amplitude. We will keep the external legs in the 
nonabelian
part of the theory.

The diagrams are then the same as in the commutative case,
with the replacement
\bea
f^{abc}\rightarrow sin(\tilde{p}^{(1)}p^{(2)}) (\delta^{ab}\delta^{c0}+
\delta^{ac}\delta^{b0}+\delta^{bc}\delta^{a0})
\eea
The final answer is then obtained from the commutative theory
by the replacement
\bea
\int {d^4P\over (2\pi)^4}{1\over (P^2-M^2)^2}C_2(G)\rightarrow \int {d^4P\over (2\pi)^4}{1\over (P^2-M^2)^2}{1\over N} sin^2(\tilde{P}q)
\eea

This leads to further IR divergences, which are however supressed in the large $N$ expansion.

\section{Orbifold Constructions}

 Constructions of SUSY gauge theories as the worldvolume theories of Dirichlet branes transverse to an orbifold singularity was pioneered in \cite{dm}. 
 The orbifold action is accompanied by an action of the discrete point group $\Gamma$ on the Chan-paton matrices, via a finite dimensional matrix representation. In \cite{dm} the regular representation was utilized. 
 This was further developed by the introduction of fractional branes \cite{fractional}, corresponding to Chan-Paton factors in  arbitrary representations of $\Gamma$.

 This construction allows for a  construction of a general class of gauge theories, with a prescribed matter content and interactions.
As the methods involved are well known, we refer the reader to \cite{dm}  for a more detailed derivation of results  used below. 

We are interested in putting D3 branes transeverse to an orbifold. We choose for simplicity two types of orbifolds:

 First, we can consider orbifolds of the form
 $C^2/Z_m$, when interested in $\mathcal{N}$=2  SUSY theories. The single generator of the orbifold  acts on two complex coordinates $X_1, X_2$ as follows:
\bea
& X_ 1 \rightarrow e^{2\pi i  /m} X_1 \nonumber \\
& X_2 \rightarrow e^{-2\pi i/m} X_2
\eea

One obtaines
a product gauge theory $U(N_1)\times ... \times U(N_m)$, with a bi-fundamental 
hypermultiplet, $(N_r, \bar{N}_{r+1})$, for each neighboring gauge groups
 (which are cyclically ordered). The case where all the factors $N_r$ are identical is the case studied in \cite{dm}. In this case the theory
turns out to be conformal. This is reflected in the orbifold model having 
no twisted sector tadpoles \cite{lr}. For other choices of integers $N_r$, one can have logarithmic  divergences, for example a non-vanishing $\beta$ function.
 This class of gauge theories has $\mathcal{N}$ =2  supersymmetry, and are therefore non-chiral.

 Furthermore, we can consider orbifolds of the form
 $C^3 /(Z_m \times Z_n)$, when interested in  $\mathcal{N}$=1 supersymmetric theories.
 The $\mathcal{N}$ =2  example is a special case of this class of orbifolds.
The two generators of $\Gamma$, $\alpha, \beta$, act as above on the complex planes spanned by 
$(X_1, X_2)$ (for $\alpha$), and on the plane spanned by $(X_2, X_3)$(for $\beta$). The action of a group element $\alpha^k \beta^{k'}$ on the Chan-Paton factors is given by a matrix $\gamma_{(k,k')}$.

The matter content can be summarized by the brane box rules \cite{hz}.
 The gauge group is a product of unitary gauge group, one for each number
$N_{rs}$, where $r=1,...m$ and $s=1,...,n$. There are also  chiral multiplets in  bifundamental representations of neighbouring gauge groups. If we cyclically order $r,s$, there are the following chiral multiplets for each $r,s$:

\bea
&(N_{r,s}, \bar{N}_{r-1,s}),(N_{r,s}, \bar{N}_{r,s+1}),(N_{r,s}, \bar{N}_{r+1,s-1})
& \mbox{in fundamental of} \, U(N_{r,s}) \nonumber\\
&(N_{r+1,s}, \bar{N}_{r,s}),(N_{r,s-1}, \bar{N}_{r,s}),(N_{r-1,s+1}, \bar{N}_{r,s}) 
&\mbox{in the anti-fundamental} 
\nonumber \\
&
\eea

This orbifold  theory can be chiral, and care has to be taken to obtain anomaly free gauge theories. This is done by cancelling a certain class of tadpoles      \cite{lr}.
 These are dubbed dimension zero tadpoles in \cite{lr}.  They are tadpoles
of (unphysical) twisted sector fields which are allowed to propagate in a dimension zero plane (a point) in $C^3$.  Cancellation of such tadpoles is a consistency condition which has to be imposed in orbifold theories \cite{gp}.
 This means that the quiver gauge theory is consistent if and only if the 
complete string theory is consistent in the corresponding background.

 Still, fairly general gauge theories can be obtained by this construction, consistent with gauge anomaly cancellation. The logarithmic divergences of those theories result from 
  tadpoles for closed string fields which were dubbed ''partially twisted'' in  \cite{lr}. Those are allowed tadpoles for  twisted sector fields which  propagate in a (real) dimension 2 plane in $C^3$. We call those twisted sector fields dimension 2 fields in what follows.

To describe the relation more precisely, denote the beta function coefficients of each non-abelian gauge group
factor by $\beta_{r,s}$. These coefficients are given by:
\beq
\beta_{rs}= 3 N_{r,s} - \frac{1}{2} \left(N_{r-1,s} + N_{r,s+1} +
N_{r+1,s-1} + N_{r+1,s} + N_{r,s-1} + N_{r-1,s+1} \right)
\eeq

 In the orbifold description, there  are $mn$ twisted sector scalar fields, denoted by
$\chi_{k,k'}$. These are  fields twisted by the generator $\alpha^k \beta^{k'}$ of the orbifold group. Some of
 those twisted sector scalars, for example those which are twisted only by one of the factors in $Z_m \times Z_n$,  propagate in dimension two plane in $C^3$.
The tadpoles for those fields encode the beta function coefficients \cite{lr}. We review this correspondence and compare it to the noncommutative case in the next section.

  It is now straightforward to  construct the noncommutative version of the quiver gauge theories. 
 In all of the models described above , the spectrum includes an untwisted NS-NS two form. 
 Therefore one can use  
the Seiberg-Witten construction \cite{sw}, and obtain  non-commutative
quiver  gauge theories with the matter content described above. We study this realization of the gauge theories below.

\section{Closed String Modes}

  Having broken the  $\mathcal{N}$=4 supersymmetry, which exists on D3-branes in flat space, we expect to discover the phenomena discussed in \cite{msv}. In particular, 
the logarithmic UV divergences  in SUSY gauge theories are now accompanied by
logarithmic IR singularities.  As reviewed above, one then discovers  closed string modes when trying to reproduce the correct logarithmic singularity
within a Wilsonian effective action.

 In the present context, having obtained the gauge theory from a string theory,
one should be able to account explicitly for the required closed string modes.
 A closed string mode $\chi$ is introduced for every case there is a logarithmic UV divergence in the commutative limit. The field $\chi$ couples linearly to a relevant or marginal operator in the gauge theory, and is allowed to propagate in two 
dimensions transverse to the brane.

 In particular, for the quiver $\mathcal{N}$ =1 gauge theories, there are 
$mn$ independent $\beta$-function coefficients. As shown above, the Wilsonian effective 
action  is then forced to have additional 
''closed string`` fields $\chi_{rs}$, one for each gauge factor $U(N_{rs})$.
 They propagate in two dimensions transverse to the brane, and couple linearly to the operator $Tr(F^2)$  in each gauge factor. Note that the latter coupling is not gauge invariant, as the pure gauge theory has no local gauge invariant operators. Presumably, the linear coupling is a part of an infinite series
of terms that couple $\chi_{r,s}$ to the gauge theory. Only the leading order 
term in such series contributes to the logarithmic divergence\footnote{Similarly, the divergence is not sensitive to the difference between regular product and *-product, when multiplying $\chi$ with the corresponding operator.}.

 In order to gain intuition about the fields $\chi_{r,s}$ and their couplings
we 
consider again the commutative quiver theories.
Consider calculating the $\beta$-functions  in open string theory. This 
can be extracted from the two point function of the gauge fields,  $\left<A^{(1)} A^{(2)}\right>$,
on the annulus.
 Indeed, in the limit when the annulus degenerates to a loop of open string modes, the 
 annulus reduces to the standard gauge field self-energy diagram. The existence of a non-vanishing beta function manifests itself in a logarithmic divergence
in the integration over  the Schwinger parameter, which is the modulus of the annulus. The relevant part of the diagram is then:

\beq
\label{beta}
A = v_4 \int \frac{dl}{l} \sum_{r,s}
 \beta_{r,s} \, Tr(F_{rs}^2) 
\eeq

where $v_4$ is the volume of the non-compact directions in string units, and $F_{r,s}$ is the field strength of the gauge group $U(N_{r,s})$. The Schwinger parameter represented by the modulus of the annulus is denoted by $l$

 Now, one can evaluate the annulus diagram in the closed string channel, 
where it reduces to an exchange of closed string modes. The only contribution to a logarithmic modular divergence was shown to arise 
from the dimension 2 twisted sector fields.  Concentrating on the contribution of
those fields to the annulus diagram reproduces the logarithmically divergent part of the self-energy diagram, equation (\ref{beta}). On the other hand it can be written as:

\beq
A= \sum_{k,k'} Tr(\gamma_{(k,k')})  Tr (\gamma_{(k,k')}\lambda_a^{rs} \lambda_b^{r's'} )  F^a_{rs} F^b_{r's'}\int \frac{dt}{t}
\eeq

Here $ \lambda_a^{r,s} , \lambda_b^{r',s'}$ are the Chan-Paton matrices of the two gauge fields. The closed string modulus is $t = \frac{1}{2l}$. The sum is constrained to include 
only dimension 2, or partially twisted, sector fields.

 The amplitude
therefore factorizes:

\beq
A= \left< A^{(1)} A^{(2)} \chi_{(k,k')} \right> \int \frac{dt}{t} \left<
\chi_{(k,k`)} \right>
\eeq

The logarithmic divergence $\int \frac{dt}{t}$ comes from a massless 
closed string field propagating in two transverse dimensions.  We note that this is the correct  factorization of the diagram where both open string vertex operators are on the same boundary component of the annulus.  The  diagram which has
the vertex operators on different boundaries factorizes differently, but does not contributes to the self-energy of the non-abelian gauge bosons. 

The argument above gives the following relations between the linear couplings
of the fields $\chi_{(k,k')}$ and the $\beta$-function coefficients:

\beq
\label{quad}
\sum_{k,k'} Tr(\gamma_{(k,k')}) 
Tr (\gamma_{(k,k')} \lambda_a^{rs} \lambda_b^{r's'} ) = \beta_{r,s} \delta^{ab}
\eeq
where the sum is again over the partially twisted sectors only. The first
factor on the left hand side is the tadpole of $\chi_{(k,k')}$ and the second factor
is the coupling of $\chi_{(k,k')}$ to the operator $Tr(F^a_{rs} F^b_{r's'})$ in the gauge theory.

 We see that the non-abelian $\beta$-functions translate in the closed string channel to the existence of linear couplings between closed string fields $\chi_{(k,k')}$ and the operators $Tr(F^2)$ and ${\cal 1}$  of the gauge theory.
This is very similar to the couplings needed in the non-commutative case \cite{msv}. In the commutative case the contribution of those closed string modes to open string scattering
 vanishes in the decoupling limit, as explained below.

 We now turn to the non-commuatative case and discuss the effects of the fields $\chi_{(k,k')}$.
The  Wilsonian effective action of the quiver gauge theory is forced to have 
some  fields $\chi$ which couple to operators in the gauge theory. We see that in
  the orbifold models there are natural candidates for the fields $\chi$. As they are predicted to propagate only in two extra dimensions they must be 
the partially twisted sector fields discussed above. Furthermore,  we saw that
 those closed string fields have linear couplings to operators in the
gauge theory.  

 To compare explicitely to the prescription of \cite{msv}, we consider the couplings of the fields $\chi_{(k,k')}$. First, their kinetic terms live in 6 dimensions, and couple to the closed string metric. Therefore their bulk action
is:
\beq
I_{bulk} = \int d^6 x \left[ \partial_\mu \chi \partial_\nu \chi g^{\mu\nu}
+ \partial_a \chi \partial_b \chi g^{ab} \right]
\eeq

where $\mu,\nu$ are the commutative directions, on and off the brane, and 
$a,b$ are the non-commuting coordinates. The closed string metric in the non
commuting directions is given in the Seiberg-Witten limit as:
\beq
G_{ab} = \frac{{\alpha'} ^2}{(\Theta^2)^{ab}}
\eeq

There are also linear couplings to the operators $Tr(F^2_{rs})$ and $1$
in the gauge theory, as discussed above\footnote{The linear couplings can be calculated from disc diagrams in string theory. The presence of a $B$-field 
merely changes the commmutative gauge fields to non-commutative ones}.

   In \cite{msv}  we are instructed to include the closed string modes
 $\chi$ up to a certain momentum scale, $\frac{1}{\alpha' \Lambda}$,
in the transverse directions.  This scale is clear if we consider open-closed channel duality. The effect of integrating out open strings of momenta higher
than $\Lambda$ can be summarized by inclusion of closed strings up to 
momenta $\frac{1}{\alpha' \Lambda}$. We comment on this cutoff further below.

Finally, the relation (\ref{quad}) is exactly  the correct
relation between the linear couplings of the fields $\chi_{(k,k')}$ and
the IR divergences. Such relation guarentees that  closed string mode exchange
 diagrams indeed reproduce all the IR divergences, to the leading order in the 'tHooft large $N_c$ expansion. 

The crucial difference between the commmutative and noncommutative cases lies in the relation between the open and closed string metrics, giving a different  scaling of the closed string metric with $\alpha'$. 
The Seiberg-Witten decoupling limit sends $\alpha'$ to zero, while keeping open string quantities fixed. This includes the UV cut-off scale $\Lambda$.

 In the commutative case 
the effect of the fields $\chi$ vanishes in the decoupling limit $\alpha' \rightarrow 0$, up to 
counterterms which renormalize the Wilsonian effective action.  This fits with the interpretation of these effects as summarizing high momentum 
gauge theory effects. 

In the noncommutative case, as shown in $\cite{msv}$, the effects of the closed string modes are non-vanishing in the decoupling limit.   Integrating out 
the gauge theory high momentum modes does not decouple them in the
usual way. Rather, they have some IR effects which are nicely summarized
 by inclusion of some light closed string modes.

\section{Conclusions}

 To summarize, we find that  closed string modes, conjectured by \cite{msv}
to be a necessary ingredient of the Wilsonian effective action, do indeed exist
in a concrete example. In the example studied, working with a finite  cutoff
couples the open string back to the closed string in  a controllable manner.
 Using the closed string fields we were able to reproduce the infrared behaviour in the leading order in the large $N_c$ limit.
 
 The discrepancy has to do with the overall $U(1)$ contribution in the loop, and is independent of the details of the gauge group or the matter content. This suggests an additional ''singleton`` $\chi$
field to account for this divergence. It would be interesting to discover that closed string directly.

Another puzzling aspect of the analysis is the origin of the cutoff on the 
closed string momenta.  The effect of integrating out open strings of momenta higher
than $\Lambda$ is summarized by inclusion of the massless closed strings, which
have  
momenta up to $\frac{1}{\alpha' \Lambda}$. 
 However, this momentum scale  is higher than the 
string scale, and therefore it is not clear why massive closed string modes 
should  not be  included\footnote{We thank M. Berkooz for conversations on the subject.}.   This might have to do with supersymmetry, as suggested in \cite{steve}. We note that any such contribution is surprising, and  it is conceivable that only a small subset of closed string mode can produce the correct behaviour. For example, massless closed string states that propagate in 
more than two dimensions do not contribute to the infrared divergences. We do not, however, have a clear understanding why massive twisted sector mode apparently make no contribution.

Finally, 
in the context of orbifolds, the quadratic divergences present a puzzle. The 
results in the supersymmetric cases discussed above strongly suggest that 
quadratic divergences can be accounted for by dimension zero tadpoles: tadpoles for twisted sector scalars that are forced to a point in the transverse space.
 However, consistency condition for string on orbifolds require vanishing of all such tadpoles \cite{gp}.  These conditions stem from an inconsistent equation of motion for such fields, and naively have nothing to do with supersymmetry.

\section{Acknowledgements}
 We thank O. Aharony, M. Berkooz, M. Douglas, H. Liu, and G. Moore
for useful conversations.

The research was supported in part by DOE grant DE-FG02-96ER40959.

\end{document}